\documentclass[sigconf]{acmart}
\usepackage{diagbox}

\AtBeginDocument{%
  }

\setcopyright{acmcopyright}
\copyrightyear{2023}
\acmYear{2023}
\acmDOI{XXXXXXX.XXXXXXX}

\acmConference[MM '23]{Proceedings of the 31th ACM International Conference on Multimedia}{Oct 29--Nov 03, 2023}{Canada, Ottawa}
\acmBooktitle{Proceedings of the 31th ACM International Conference on Multimedia (MM '23), October 10--14, 2023, Canada, Ottawa}
\acmPrice{15.00}
\acmISBN{978-1-4503-XXXX-X/18/06}

\acmPrice{15.00}
\acmISBN{978-1-4503-XXXX-X/18/06}

\begin{document}

\title{RetinexFlow for CT metal artifact reduction}

\author{Jiandong Su}
\email{jd.su@siat.ac.cn}
\authornotemark[1]
\affiliation{%
  \institution{Shenzhen Insititute of Advanced Technology, CAS}
  \streetaddress{P.O. Box 1212}
  \city{Shenzhen}
  \state{Guangdong}
  \country{China}
  \postcode{518034}
}

\author{Ce Wang}
\email{wangce@ict.ac.cn}
\authornotemark[1]
\affiliation{%
  \institution{Institute of Computing Technology, CAS}
  \streetaddress{P.O. Box 1212}
  \city{Beijing}
  \state{Beijing}
  \country{China}
  \postcode{}
}

\author{Yinsheng Li}
\email{ys.li@siat.ac.cn}
\affiliation{%
  \institution{Shenzhen Insititute of Advanced Technology, CAS}
  \streetaddress{P.O. Box 1212}
  \city{Shenzhen}
  \state{Guangdong}
  \country{China}
  \postcode{518034}
}

\author{Kun Shang}
\email{Kun.Shang@siat.ac.cn}
\affiliation{%
  \institution{Shenzhen Insititute of Advanced Technology, CAS}
  \streetaddress{P.O. Box 1212}
  \city{Shenzhen}
  \state{Guangdong}
  \country{China}
  \postcode{518034}
}
\author{Dong Liang}
\email{dong.liang@siat.ac.cn}
\affiliation{%
  \institution{Shenzhen Insititute of Advanced Technology, CAS}
  \streetaddress{P.O. Box 1212}
  \city{Shenzhen}
  \state{Guangdong}
  \country{China}
  \postcode{518034}
}

\renewcommand{\shortauthors}{Su et al.}

\begin{abstract}
Metal artifacts is a major challenge in computed tomography (CT) imaging, significantly degrading image quality and making accurate diagnosis difficult. However, previous methods either require prior knowledge of the location of metal implants, or have modeling deviations with the mechanism of artifact formation, which limits the ability to obtain high-quality CT images. In this work, we formulate metal artifacts reduction problem as a combination of decomposition and completion tasks. And we propose RetinexFlow, which is a novel end-to-end image domain model based on Retinex theory and conditional normalizing flow, to solve it. Specifically, we first design a feature decomposition encoder for decomposing the metal implant component and inherent component, and extracting the inherent feature. Then, it uses a feature-to-image flow module to complete the metal artifact-free CT image step by step through a series of invertible transformations. These designs are incorporated in our model with a coarse-to-fine strategy, enabling it to achieve superior performance. The experimental results on on simulation and clinical datasets show our method achieves better quantitative and qualitative results, exhibiting better visual performance in artifact removal and image fidelity.
\end{abstract}


\maketitle

 \section{Introduction} 
Computed tomography (CT) is an indispensable imaging technology that assists clinical decision-making for medical diagnosis and treatment with high-quality anatomical representations of human body. However, metallic implants inserted into the patient's body, such as dental fillings and hip prostheses, would lead to corrupted information in X-ray projections (sinograms) and cause undesirable star-shape or streak artifacts in the reconstructed CT images~\cite{de1998metal}. These artifacts not only destroy the observation of anatomical details for affect the clinical diagnosis, but also make dose calculation problematic in radiation therapy to limit the diagnostic value of CT scans~\cite{kalender1987reduction}. With the widely use of metallic implants, how to reduce metal artifacts has become an important problem, which gains increasing attention from the CT community.

Numerous metal artifact reduction (MAR) methods have been proposed in the past decades. Since the metal artifacts in CT images have non-local, structured characteristics, the previous MAR methods~\cite{hsieh2000iterative,kachelriess2001generalized,park2016metal} mainly focus on the sinogram domain by modeling the physical effects of the presence of high atomic number metals. However, the metal trace regions in sinogram domain are often severely corrupted such that these methods are limited in achieving satisfactory results. The other perspective regards the metal trace regions as the missing areas and fills them by linearly interpolating with the adjacent unaffected projection views~\cite{kalender1987reduction,prell2009novel,muller2009spurious,meyer2010normalized}. As these methods cannot accurately recover the metal trace information, the inconsistency between interpolated values and the unaffected values often causes secondary artifacts in CT images. In addition, several works~\cite{zhang2013hybrid,wang2013metal} recover the affected sinogram by estimating the prior information of various tissue from the  other uncorrupted image. Recently, some deep learning methods~\cite{park2018ct,zhang2018convolutional,yu2020deep} have proposed to directly learn a mapping function from the sinogram domain to the artifact-reduced image. We call these  methods working in sinogram domain as methods based on sinogram-domain enhancement (SE). Despite the success achieved by the above SE methods, there are still significant limitations due to the requirement of metal trajectories and the newly generated artifacts in reconstructed CT images. In practice, it is difficult to obtain the location of metal implants, and the additionally secondary artifacts makes the MAR uneffective. 

Meanwhile, some researchers~\cite{mehranian2013x,huang2018metal,wang2018conditional,liao2019generative,liao2020adn,lyu2020encoding,wang2021dicdnet} consider MAR as an image-domain restoration (IR) problem, and reduce the metal artifacts with image-to-image translation networks, which no longer rely on the position information of metals. For example, Huang et al.~\cite{huang2018metal} introduce deep residual learning to reduce metal artifacts in cervical CT images. Wang et al.~\cite{wang2018conditional} propose to use the conditional generative adversarial network (cGAN)~\cite{isola2017image} to reduce metal artifacts in CT images. Then, Liao et al.~\cite{liao2020adn} introduce a artifact disentanglement network that disentangles the metal artifacts from CT images in the latent space by unsupervised learning. Recently, Lin et al.~\cite{lin2019dudonet} develop a dual-domain learning method to improve the performance of MAR by involving sinogram enhancement as a procedure. Wang et al.~\cite{wang2021dicdnet} propose an deep interpretable convolutional dictionary Network for MAR, which uses LI~\cite{kalender1987reduction} to enhance sinogram by considering the non-local repetitive streaking priors of metal artifacts. As there is no physical priors to regularize the models, there exist unreasonable patterns of anatomical structures and image contrast in the recovery, which limits the usage of image domain methods in real clinical scenarios.

In this work, we propose a novel image-domain method, named RetinexFlow, for reducing metal artifacts in reconstructed CT images. Different from the previous SE/IR methods, we formulate MAR as a combination of decomposition and completion tasks. For decomposition task, we design a feature decomposition encoder by Retinex theory~\cite{land1971lightness,land1977retinex}, which bridges the physical prior modeling missing in previous works. It decomposes the variant component and inherent component in a CT image, and then extracts the inherent feature. For completion task, we convert it to a distribution transformation task, and design a conditional feature-to-image flow module to complete the metal
artifact- free CT image step by step through a series of invertible transformations. Since the transformation is working at the distribution level, it does not depend the information of metal implants for every image to assist the artifact removal, greatly reducing the interaction complexity. 

To sum up, our contributions are as follows:%需改
\begin{itemize}
\item{We formulate MAR as a combination of decomposition and completion tasks. To avoid two-stage training for decomposition and completion separately, we propose an end-to-end conditional learning framework in a coarse-to-fine way.}
\item{Inspired by the Retinex theory, the CT image is decomposed to inherent component and variant component. We set the feature decomposition encoder for coarse extracting the inherent feature of the object itself in the CT image.}
\item{We further use normalizing flow to refine in feature space, which rather than directly processing the image or sinogram domains. It progressively narrows down the solution space, resulting in a cleanest solution. }
\item {The quantitative and qualitative results on the simulated DeepLesion dataset demonstrate that RetinexFlow is capable to remove artifacts while preserving anatomy details. Moreover, when testing on the clinical CT pelvic dataset from different anatomies, our method shows better generalization performance and is effective in removing artifacts.}
\end{itemize}

\section{Background and Motivation}%需改
\subsection{Problem formulation}
Different human body tissues have different X-ray attenuation coefficients $\mu$. Considering the 2D CT image, we use $X = \mu(x, y)$ to represent the anatomy structure. According to the Lambert-Beer law~\cite{beer1852bestimmung}, with a polychromatic X-ray source, sinograms $Y$ of anatomy structures are determined by the following model with energy distribution $\eta(E)$:
\begin{equation}
Y = -log \int{\eta(E)\exp\{-PX(E)\}dE},
\label{eq:lb raw}
\end{equation}
where $P$ denotes the forward projection operator. In practical CT imaging, we recover the 2D image $X(E)$ from the measured sinograms $Y$. Normally, without metal implants, $X(E)$ is approximately equal to a constant relative to the X-ray energy $E$, and therefore, $X=X(E)$. Then reconstruction $X^{\dag}$ can be inferred with various imaging algorithms $P^{\dag}$~\cite{chen2017low, wang2021improving}. When there exists the metal implants $M(E)$, $X(E)$ would suffer from large variations and $X = X(E) + M(E)$. Thus, the equation~(\ref{eq:lb raw}) becomes:
\begin{equation*}
Y =  - log \int{\eta(E)\exp\{PX(E) -PM(E)\}dE}.
\label{eq:metal insert}
\end{equation*}
In this case, when we still back-project the corrupted sinograms $ \title{Y} $ and reconstruct with $P^{\dag} $, this results
\begin{equation}\label{recon}
X_{M}^{\dag} = P^{\dag} Y = X^{\dag} - P^{\dag} lo g\int{\eta(E) \exp\{ -PM(E) \} }dE.
\end{equation}
Thus, the reconstruction error between $X_{M}^{\dag}$ and $X$ is formulated as
\begin{equation*}
e_{M} = X_{M}^{\dag} - X.
\label{recon loss}
\end{equation*}

\begin{figure}[t!]
  \includegraphics[width=8.5cm]{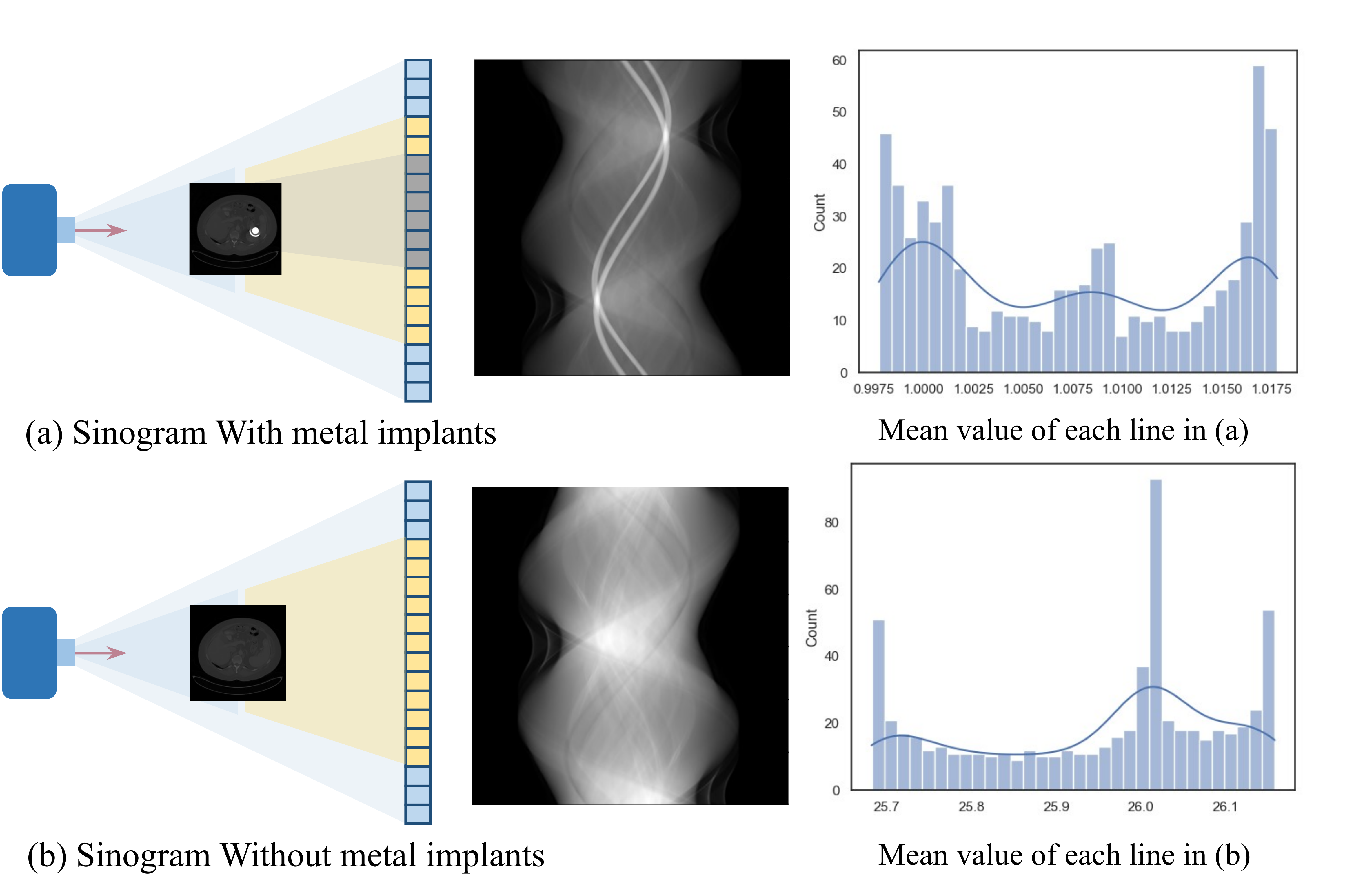}
  \caption{
  Under the same imaging setting, we analyze the impact of the presence or absence of metal implants on the imaging process. Targeting this, we statistically count mean value of each line in obtained projection data w/wo metal accordingly in subfigures (a) and (b), and the histograms show that the presence of metal implants has a significant impact on the projection values obtained by the detector, and brings significant attenuation. Such inconsistency between two histograms make it hard to be recovered with previous imaging algorithms. }
  \label{fig:numer}
\end{figure}

\noindent If the projection data is consistent, then $e_{M}$ should be close to $0$ over the entire object area. However, when projection data is corrupted, we will not be able to obtain accurate reconstruction results. Clearly, at this time, $e_{M}$ will have a significant deviation over the entire object area. As the counted mean value of each line in obtained sinograms w/wo metals in Fig.~\ref{fig:numer}, when there are metal implants in the human body, the number of photons reaching the detector is greatly attenuated. Similar, the numerical values of the sinograms will undergo significant attenuation, manifested as streak artifacts or black shadows in the image.

\subsection{Motivation}
To minimize $e_{M}$, SE methods fill the metal-corrupted areas with estimated values, while the required prior knowledge of the metal trace makes it practically ineffective. Besides, directly correcting the sinograms, especially the extra peak shown in Fig.~\ref{fig:numer}, across metals of different size and shape is difficult. The IR methods formulate the MAR problem as an image restoration problem. Such methods define the linear relationship between image content and metal artifacts, and use disentanglement procedure to separate the two components to obtain a corrected image. However, there exist unreasonable patterns of anatomical structures and image contrast in the Restored image without using physical priors. Some researchers have attempted to combine both domains by introducing sinogram enhancement in the image domain to improve network performance. However, these methods still directly use the image domain improved output as the final reconstructed image, which may result in anatomical structure changes in the output image.  

\begin{figure}[ht]
\includegraphics[width=8cm]{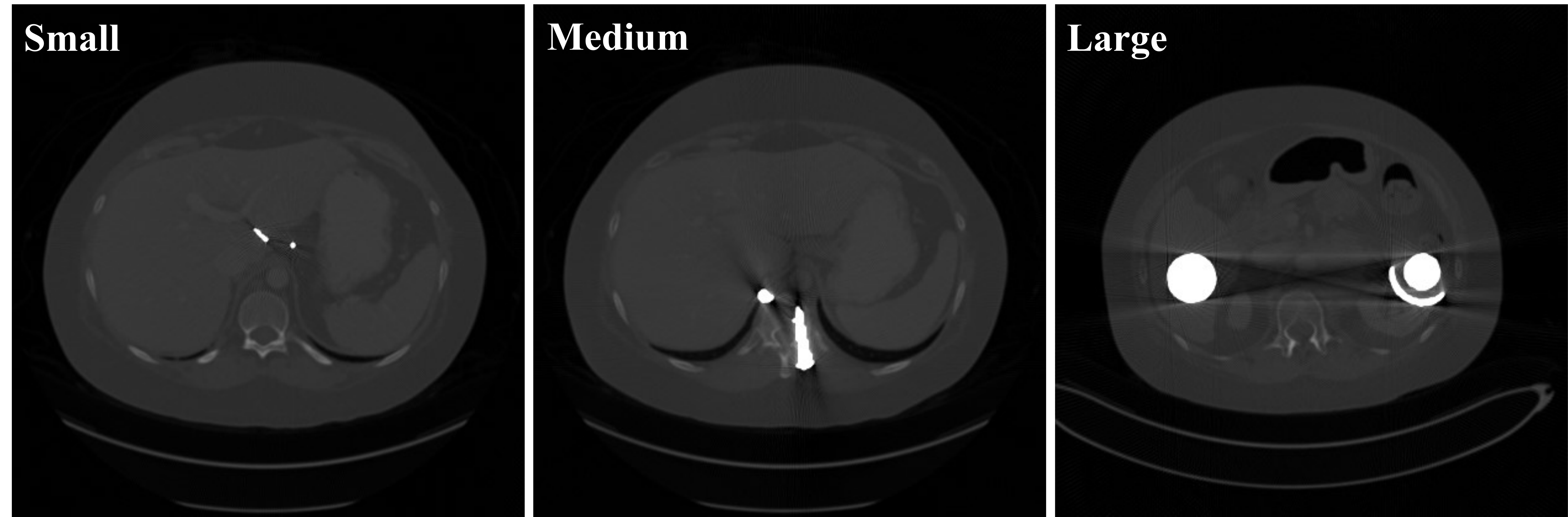}
\caption{Some examples about the different "light source" simulated in the same image from DeepLesion dataset.}
\label{fig:metal_implants}
\end{figure}

In addition, \textbf{Notice that anatomy structure of CT image content is much less than natural image content, making it easier to overfit to special size or shape of metal.} And the presence of black shadow areas in the image, especially when facing with large or irregular metal-implants (some examples are shown in Fig.~\ref{fig:metal_implants}), it is difficult to solve the missing value filling problem.

All above discussions motivate us to find a new solution for MAR. We regard MAR as a combination of decomposition and completion tasks, and further model it in image domain without depending on sinogram.

\section{method}
\begin{figure*}[ht]
\includegraphics[width=\textwidth]{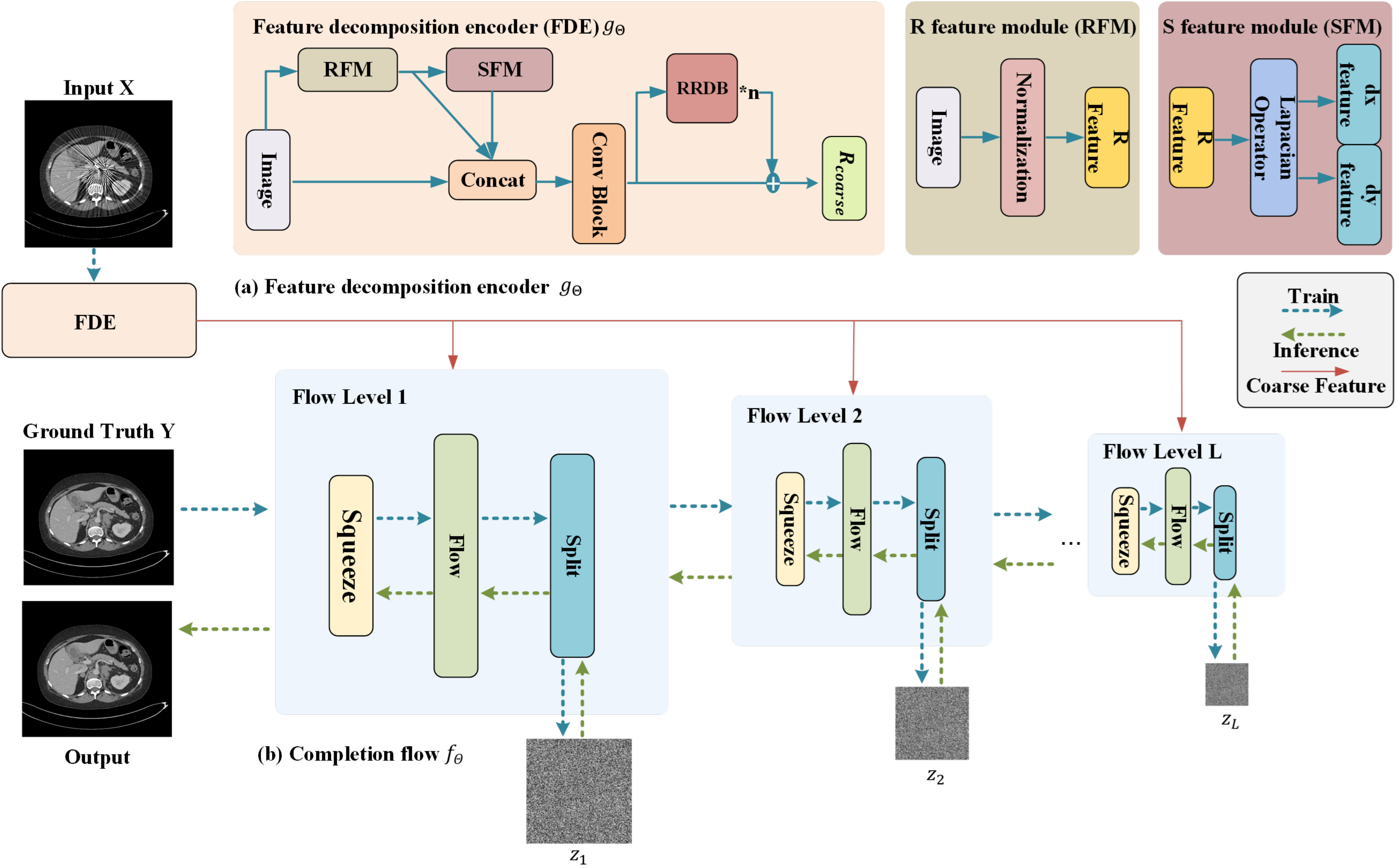}
\caption{An illustration of the RetinexFlow architecture. RetinexFlow consists of an feature decomposition encoder $g_\theta$ and an Completion flow $f_\theta$. 
%Firstly we use $g_\theta$ reduce the search of the feasible artifact-free solution space.Given an coarse estimated of $R_{coarse}= g_{\theta}(X)$ about the metal artifact-free component, the $f_\theta$ operates at multiple scale levels and processes the input through a series of reversible transformations. 
}
\label{fig:framework}
\end{figure*}
Consider MAR problem in image domain, as show in Fig.~\ref{fig:metal_implants}, different metal implants present different white spots with different metal artifacts in the reconstructed CT image. Inspired by Retinex Theory~\cite{land1971lightness,land1977retinex}, which assume that images under the same scene with different light conditions can have different illumination components, while share the same reflectance component. we regard the metal implant as a "light source" in a CT image. And a metal artifact CT image $X$ is formulated as:
\begin{equation*}
    X = L \odot R,
\end{equation*}
where $L$ represents the illumination component decided by metal implants, $R$ represents the reflection component of the inherent properties of the object itself, and $\odot$ means the element-wise product. 

If such a decomposition for the CT image is accurately estimated, the corresponding metal artifact-free image can be obtained. However, the estimated metal artifact-free component $R_{est}$ may contain unexpected degradations, such as noise and contrast biases, by existing method~\cite{fan2020integrating}. For further improving the estimation $R_{est}$, we propose RetinexFlow model, which is shown in Fig.~\ref{fig:framework}, for obtain the cleanest solution with the greatest conditional probability in a coarse-to-fine way. It contains two main modules, the first one is for decomposing $L$ and $R$, and give a coarse estimation $R_{coarse}$. Then, we use a conditional normalizing flow to refine $R_{coarse}$ for obtain the cleanest metal artifact-free image $Y_{est}$.

\subsection{Feature decomposition encoder}
We design a preliminary feature decomposition extractor (FDE), which will reduce the search of the feasible artifact-free solution space. First, we take an normalization operator on input (metal artifact) image $X$ for suppressing the influence of metal artifacts and other noise, which is: 
\begin{equation*}
  N = \frac{dX}{\sum_{k=1}^{d}vec(X)},
\end{equation*}
where $vec(\cdot)$ means the vectorization operator, $d$ is the dimension of $vec(X)$, respectively.  Secondly, in order to preserve more details about edge and structure,  we further conduct the vertical and horizontal gradients of $N$ as feature $S$:
\begin{equation*}
    S= concat(\nabla_c N, \nabla_r N),
\end{equation*}
where $concat(\cdot) $ is a concatenation operator. Finally, the result of $concat(X, N, S)$ as the input of RRDB~\cite{wang2018esrgan}. Notice that the dimension of $concat(X, N, S)$ is larger than the preliminary input, we remove the upsampling layer in RRDB~\cite{wang2018esrgan} for keeping the same dimension, we denoted it as modified RRDB (mRRDB). Thus, the first estimated metal artifact-free component $R_{coarse}= g_{\theta}(X)$ is accomplished, and the structure of FDE is shown in Fig.~\ref{fig:framework}~(a).

\subsection{Completion flow}
Given an coarse estimated of $R_{coarse}= g_{\theta}(X)$ about the metal artifact-free component, the metal artifact-free image $X_{est}$ can get from it. To obtain the cleanest solution, the next task is completion task. We design a feature-to-image multi-scale conditional flow module, named completion flow (CF), based on Glow~\cite{kingma2018glow} and Real-NVP~\cite{dinh2017density}, to restore the metal artifact-free CT image $Y_{est}$.

Based on the change-of-variable formula, flow-based methods~\cite{dinh2015nice,dinh2017density,kingma2018glow} map the distribution of  images to a simple prior distribution, which can realize the exact conversion between the latent feature space and image space through well-designed reversible network structures. Our CF module has a multi-scale structure consisting of L level for modeling a one-to-many mapping between a feature and its feasible solution space (image), which is shown in Fig.~\ref{fig:framework}~(b). 
%IF module learns the conditional probability distribution over all feasible artifact-free solutions through likelihood-based training, which can model the distribution of input step by step through a series of invertible transformations.
Concretely, the distribution $p_{Y|R_{coarse}}(Y|R_{coarse},\theta)$ with $f_\theta$ that maps data pairs $(R_{coarse},Y)$ to latent variables $z=f_{\theta}(Y;R_{coarse})$. Since the network is reversible, we can always accurately reconstruct $Y_{est}=f_{\theta}^{-1}(Z;R_{coarse})$ from the latent variable $Z$. 

For each scale, for facilitating information exchange along the channel dimension, an squeezing layer is used to compress the input along the channel axis and spatial dimensions. Each level performs operations in series, and a single scale operation consists of K reversible flow steps, which perform more refined reasoning. Concretely , a flow step consists of three components: \textbf{actnorm}, \textbf{$1 \times 1$ invertible convolution}, and \textbf{coupling layer}.  
\begin{itemize}
  \item Actnorm is to normalize the input;
  \item $1 \times 1$ convolution acts as a permutation convolution, mixing information along the channel dimension;
  \item Coupling layer introduces non-linearity, and the composition of multiple coupling layers allows for the model to have stronger representational power.
\end{itemize}

\subsection{Loss function}
Given a large number of training pairs $(X_i, Y_i)$ ($i$ is the sample number) of images w/wo metal artifacts, we define the MAR problem as the conditional probability distribution problem of learning to project metal-free images $Y$ from metal artifact images $X$ by minimizing the negative log-likelihood (NLL) of the loss function:
\begin{eqnarray*}\label{nll loss}
\mathcal{L} (\theta;X_i,Y_i) \!\!\!\!\!&=&\!\!\!\!\!-\log p_{Y_i|g_{\theta}(X_i),\theta} \nonumber\\
                         &=&\!\!\!\!\!-\log p_z(f_{\theta}(Y_i;g_{\theta}(X_i)))-\log \left| \det \frac{\partial f_{\theta} }{\partial Y_i}(Y_i;g_{\theta}(X_i))\right|,
\end{eqnarray*}
where $g_{\theta}(\cdot)$ is FE, and $\det$ represents the Jacobian determinant, illustrating the density transformation caused by the reversible network $f_\theta$.

\begin{table*}[htbp!]
\caption{Quantitative comparison results (PSNR/SSIM) on deep lesion datasets. The difference between our method and compared methods are statistically significant at $p < 0.01$}
\label{tab:SynResults}
\begin{tabular}{l|ccccc|l}
\toprule
Methods & Large metal & $\longrightarrow$ & Middle metal & $\longrightarrow $ & Small metal & Average \\
\hline 
Input                            & 24.12/0.6761 & 26.13/0.7471 & 27.75/0.7659 & 28.78/0.8076 & 28.92/0.8081 & 27.06/0.7586 \\
LI~\cite{kalender1987reduction}  & 27.21/0.8920 & 28.31/0.9185 & 29.86/0.9464 & 30.30/0.9555 & 30.57/0.9608 & 29.27/0.9347 \\   
NMAR~\cite{meyer2010normalized}  & 27.66/0.9114 & 28.81/0.9373 & 28.69/0.9465 & 30.44/0.9591 & 30.79/0.9669 & 29.48/0.9442 \\ 
ADN~\cite{liao2020adn}           & 31.27/0.9324 & 33.12/0.9415 & 35.21/0.9505 & 36.12/0.9596 & 37.12/0.9678 & 34.57/0.9503 \\
Dudonet++~\cite{lyu2020encoding} & 36.97/0.9855 & 39.34/0.9892 & 40.24/0.9911 & 42.80/0.9925 & 44.12/0.9941 & 40.64/0.9905 \\ 
DICDNet~\cite{lyu2020encoding}   & 37.19/0.9853 & 39.53/0.9908 & 42.25/0.9941 & 44.91/0.9953 & 45.27/0.9958 & 41.83/0.9923 \\ 
\textbf{Ours}                    & \textbf{42.92/0.9919} & \textbf{43.26/0.9936} & \textbf{46.20/0.9942} & \textbf{48.69/0.9967} & \textbf{49.21/0.9968} & \textbf{46.04/0.9946} \\  
\bottomrule
\end{tabular}
\end{table*}

\section{Experiment}
To verify the effectiveness and generalization of our method, we verify the proposed methods on both synthetic and clinical data. In addition, we also give enough discussion about RetinexFlow itself in this section.

\subsection{Dataset and Experimental Setup}
\textbf{Dataset.} We use the DeepLesion~\cite{yan2018deeplesion} dataset for training and verification. Specificly, we randomly select 4200 total clean CT images from the DeepLesion~\cite{yan2018deeplesion} to synthesize metal-corrupted images, where 4000 images are for training and the other 200 images are for testing. Then, following the procedures from Zhang et al.~\cite{zhang2018convolutional}, 100 metal masks of different size and shape are generated for corruption synthesis, of which 90 masks are used in training and 10 masks are used in testing. The simulation is conducted in a fan-beam geometry with 640 projections uniformly spaced between 0-360 degrees. All these CT images are of the size $416\times416$.\\
For further demonstrate the generalization and clinical value of our method, we choose the CT pelvic1K dataset~\cite{liu2021deep} for testing, which has many real metal artifact images. We extracted 230 2D CT images with metal implants from a 3D sequence, and each image has a size of $512 \times 512$.\\
\textbf{Compared methods.} We compare our method with several state-of-the-art CT MAR methods, including traditional iterative SE methods (LI~\cite{kalender1987reduction} and NMAR~\cite{meyer2010normalized}), IR method based on deep generative model (ADN~\cite{liao2020adn}), and IR methods based on dual domain learning (Dudonet++~\cite{lyu2020encoding}, DICDNet~\cite{wang2021dicdnet}). Among these methods, LI~\cite{kalender1987reduction}, NMAR~\cite{meyer2010normalized}, and DICDNet ~\cite{wang2021dicdnet} require prior information about the metal implant as a constraint, while ADN~\cite{liao2020adn}, Dudonet++~\cite{lyu2020encoding}, and our RetinexFlow only require a single CT image as input.\\
\textbf{Implementations.} 
We implement all the experiments with the Pytorch framework. During the training, models are trained for 50 epochs on a single NVIDIA A6000 GPU with a learning rate of $2\times10^{-4}$ and a batch size of 6 .We use the Adam optimizer~\cite{kingma2015adam} with parameters $(\beta1, \beta2) = (0.5, 0.999)$.  Whatever the synthetic or clinical experiment, we set flow level number $L=3$, flow-step number $K=6$, and hidden channel number $c=64$ in CF module, and freeze the $1 \times 1$ reversible convolution for stable training.\\
\textbf{Evaluation Metrics.} Images are quantitatively evaluated with the peak signal-to-noise ratio (PSNR) and strutured similarity index (SSIM)~\cite{wang2004image}.

\begin{figure*}[ht]
\centering
\includegraphics[width=0.9\textwidth]{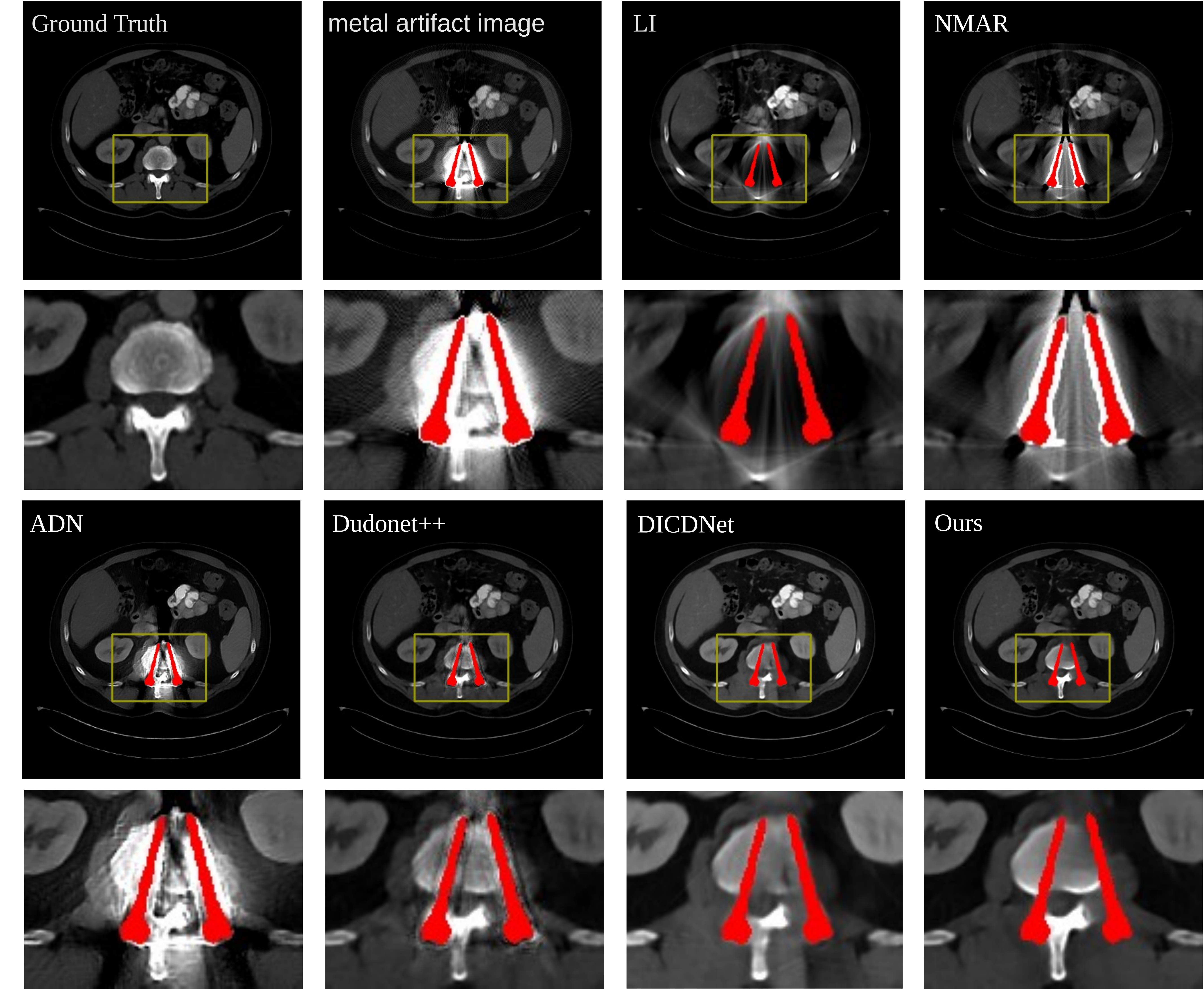}
\caption{Qualitative results with compared methods on the DeepLesion dataset. All images displayed have the same window of [-175, 275].}
\label{fig:syn}
\end{figure*}

\subsection{Results on simulation data}
\textbf{Quantitative performance}
The quantitative comparison results of our proposed method and other methods on the simulation data are shown in Table~\ref{tab:SynResults}. It is obvious that the deep-learning-based methods outperform traditional MAR methods in terms of PSNR and SSIM, indicating the superiority of data-driven methods in the MAR problem. Especially, the dual-domain-learning-based DICDNet performs better over others by incorporating LI correction results as reference values. With the introduction of Retinex theory on modeling physical prior, our method further improves the quantitative performance although we only require image-domain input and do not need metal trace as input. Specifically, We achieve a 4.21dB improvement in terms of PSNR and a little improvement in terms of SSIM over the second DICDNet, demonstrating the effectiveness of such Retinex inspired modeling in the MAR problem.

\textbf{Qualitative performance}
We further show the visual comparisons between our method and comparative methods on the simulation data in Fig.~\ref{fig:syn}. To enhance the display effect, the simulated metal implants are masked in red(All subsequent experiments are presented using this approach). Traditional method like LI~\cite{kalender1987reduction} and NMAR~\cite{meyer2010normalized} cannot accurately reconstruct the accurate anatomical structure of metal implants and the surrounding bone and soft tissues, 
while our method can achieve this. When the metal implants are relatively large, both the ADN~\cite{liao2020adn} and DuDoNet++~\cite{lyu2020encoding} fail to preserve the details of the original image well. Although DICDNet~\cite{wang2021dicdnet} takes LI~\cite{kalender1987reduction} results as inputs, the method introduces secondary artifacts which is clinically undesirable. In contrast, our method can remove most of the streaking artifacts while effectively filling the black shadow regions. Although RetinexFlow is a image-domain method, which do not utilize the LI~\cite{kalender1987reduction} results as references, the Retinex inspired learning help correct the constrast between anatomies, avoiding secondary artifacts while retaining the original structure details.

\subsection{Ablation studies}
Next, we discuss the effectiveness of several designs of our proposed method. All experiments are conducted on the simulation data to allow for quantitative and qualitative analysis.
\subsubsection{Why we use FE?} 
To illustrate why we use the designed FE as the head of IF, we first give a toy example. For comparison, a modified network without FE, which still include the mRRDB encoder, is used to demonstrate the effectiveness of FE. The quantitative results of the two methods are shown in the Table.~\ref{tab:FE}. It can be observed that the network trained using FE module performs better with 1.98dB improvement in terms of PSNR, indicating the effectiveness of the proposed FE module.

\begin{table}[t!]
\caption{Performance between Different Feature encoder}
\label{tab:FE}
\begin{tabular}{ccl}
\toprule
Method &PSNR&SSIM\\
\midrule
Ours wo FE& 44.12 &  0.9887\\
\textbf{Ours w FE} & \textbf{46.04}& \textbf{0.9946}\\
\bottomrule
\end{tabular}
\end{table}

\begin{figure}[t!]
  \includegraphics[width=8.5cm]{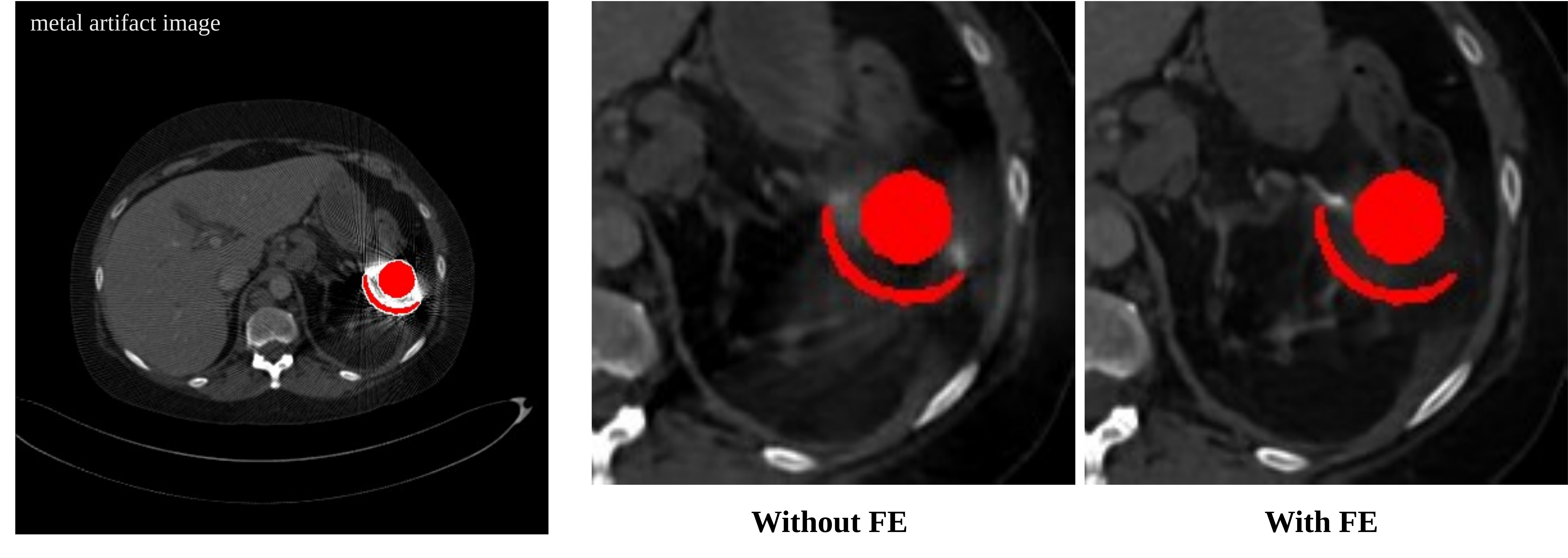}
  \caption{
  Qualitive performance between model with or without FE }
  \label{fig:FE}
\end{figure}

The corresponding visual result is shown in the Fig.~\ref{fig:FE}. It is some artifacts that remains in the CT image, and the completion in the black shadow region is not satisfied, which has been processed by  FlowNet with convolutional encoder mRRDB (without FE). Whatever quantitative or qualitative results indicate the effectiveness of FE in reducing metal artifacts.

\subsubsection{Influence of Flow-step Number}

\begin{table}[t!]
  \caption{Performance between Different Flow-steps}
  \label{tab:flow_step}
  \begin{tabular}{ccl}
    \toprule
    Flow step number&PSNR&SSIM\\
    \midrule
    1& 39.87&  0.9720\\
    3& 41.24& 0.9811\\
    \textbf{6}& \textbf{46.04}& \textbf{0.9946}\\
  \bottomrule
\end{tabular}
\end{table}

\begin{figure*}[htbp!]
  \includegraphics[width=12.5cm]{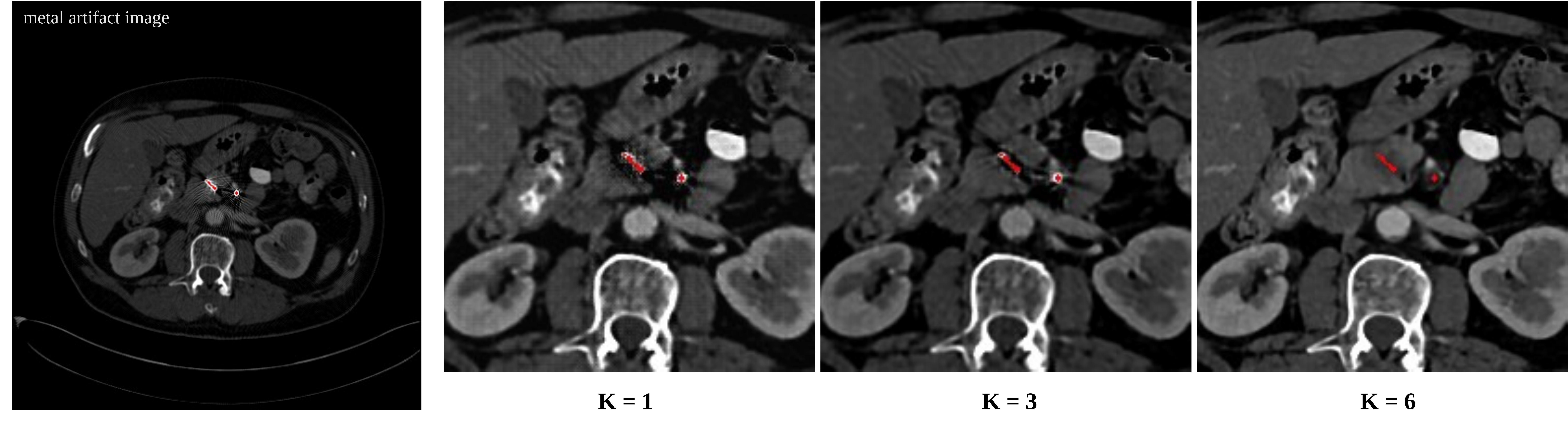}
  \caption{Qualitative performance between different flow-step number}
  \label{fig:flow_step}
\end{figure*}

CF is an important module for refining the inherent feature, and transform it to metal artifact-free image, which has multiple non-linear flow-steps. Here, we compare the impact of different numbers of flow-step in the RetinexFlow model, and show the results in Table.~\ref{tab:flow_step}. We observe that the performance of RetinexFlow consistently increases when we enlarge the number of flow-step from $1$ to $6$. Within a certain range of flow-step numbers, the higher the flow-step number, the better the quantitative performance. And the same phenomenon has also been shown in Fig.~
~\ref{fig:flow_step}, which means the CF gets deeper, the processing of features gets more refined.  

\subsubsection{Freeze $1 \times 1$ invertible convolution or not}

\begin{figure*}[t!]
\includegraphics[width=12.5cm]{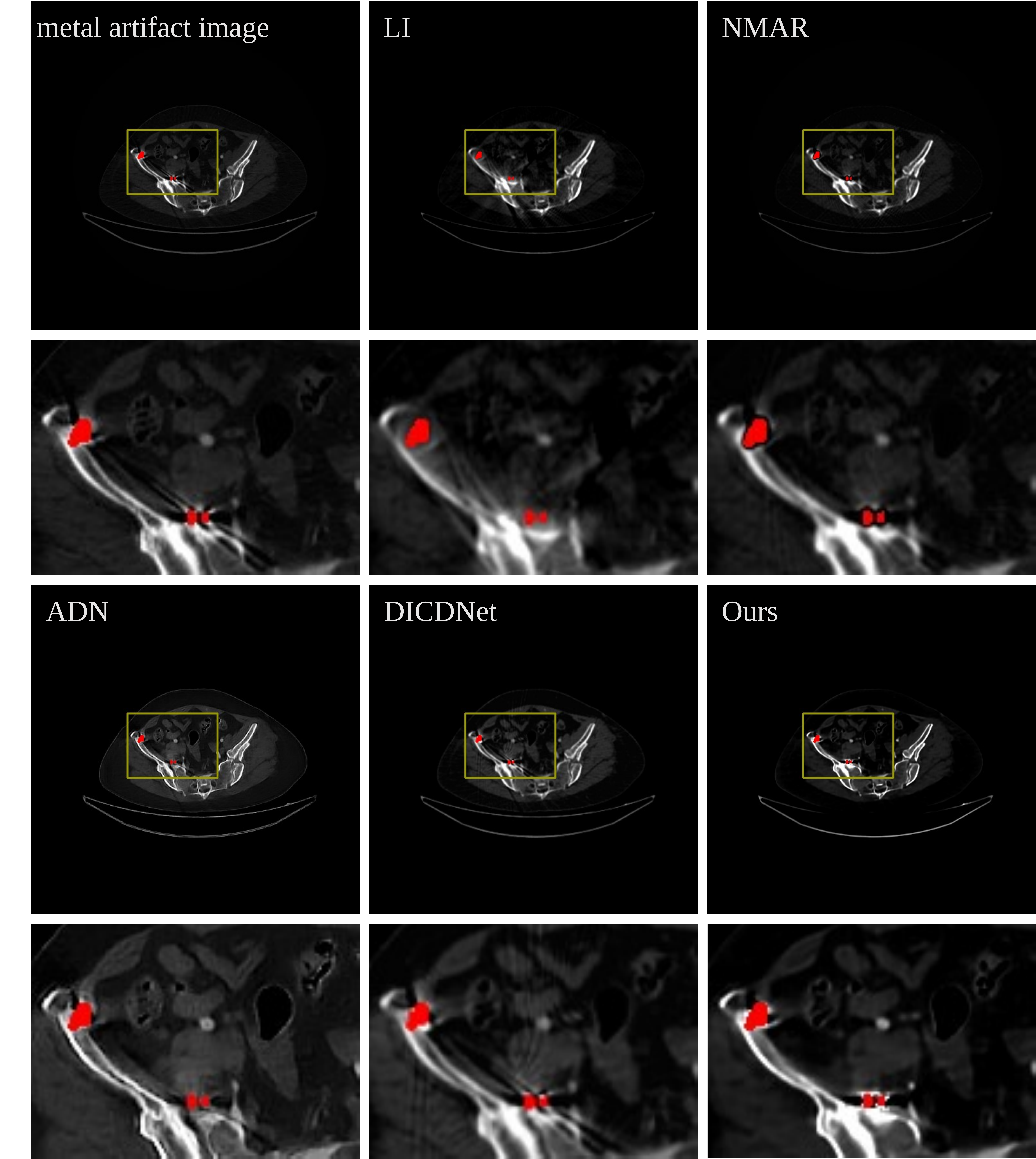}
\caption{Qualitative results with compared methods on the CT Splveic dataset.For CT images, the window width is 450 HU and the window level is 50HU. }
\label{fig:Real}
\end{figure*}

\begin{table}[htbp!]s
\caption{The experimental results of freezing $1 \times 1$ invertible convolution or not}
\label{tab:1*1}
\begin{tabular}{ccl}
 \toprule
Method &PSNR&SSIM\\
\midrule
 Our model wo freeze & 45.24& 0.9911\\
\textbf{Our model w freeze} & \textbf{46.04}& \textbf{0.9946}\\
\bottomrule
\end{tabular}
\end{table}

Based on Glow \cite{kingma2018glow}, some previous works~\cite{lugmayr2020srflow} employ the $1 \times 1$ invertible convolution layer before each affine coupling layer to mix information. However, such operation introduces instability during model training. For better use the $1 \times 1$ invertible convolution layer, We further explore whether to freeze it. As shown in Table~\ref{tab:1*1}, when we freeze the $1 \times 1$ invertible convolution layer, the model achieves 0.8dB improvement than the non-freeze one in the term of PSNR. Thus we the $1 \times 1$ invertible convolution layer, which seems not suitable for our task.

\subsubsection{Influence of hidden channels}
To ablate the model width, we train our network with different number of hidden channels in two conditional layers, which is shown in Table~\ref{tab:width}.  Decreasing the number of hidden layers leads to more artifacts in complex structures, and a larger number of channels leads to better image quilts for CT MAR. Therefore, we set the number of hidden channels as 64 in our model.
\begin{table}[ht!]
  \caption{Performance between the different hidden channel number}
  \label{tab:width}
  \begin{tabular}{ccl}
    \toprule
    Method&PSNR&SSIM\\
    \midrule
    32 & 43.21&  0.9918\\
    64 & 46.04& 0.9946\\
  \bottomrule
\end{tabular}
\end{table}

\subsection{Results on clinical data}
In order to further verify the generalization and clinical value of the proposed RetinexFlow model, we next evaluate the performance of our proposed RetinexFlow on clinical "CT pelvic1K dataset~\cite{liu2021deep}", which has many real metal artifact images. As there are no available ground truths, we only perform qualitative comparisons. Notice that all the deep-learning-based methods are trained on the DeepLesion ~\cite{yan2018deeplesion} dataset, not the clinical data "CT pelvic1K dataset~\cite{liu2021deep}". 

As shown in Fig.~\ref{fig:Real}, we find that traditional methods such as LI \cite{kalender1987reduction} and NMAR \cite{meyer2010normalized} have removed certain metal artifacts, while they introduce secondary artifacts when completing the projection domain. Among the deep-learning-based methods, ADN~\cite{liao2020adn} removes most of the streaks and dark shadows, but changes the image sharpness and still leaves residual streak-like artifacts. Although DICDNet \cite{wang2021dicdnet} can preserve the image structures well, it also introduces secondary artifacts in the final display results because of the utilization of LI. In contrast, with the Retinex inspired learning, image contrast prior is physically modeled. And the coarse to fine processing of RetinexFlow ensures that no new artifacts are introduced. Therefore, RetinexFlow removes most of the streaks and dark shadows while preserving most details of the original image.

\section{Conclusion}
In this work, we formulate metal artifacts reduction problem as a combination of decomposition and completion tasks. And we propose  a novel end-to-end image domain model based on Retinex theory and conditional normalizing flow, named RetinexFlow, to solve it. To obtain the cleanest metal-artifact-free image, the coarse to fine RetinexFlow first decomposing the metal implant component and inherent structure component, and then refines the extracted inherent feature to the cleanest metal artifact-free image. Experimental results of our experiments on simulation data indicate that the proposed method achieves the best performance, quantitatively and qualitatively. Though only using simulation data for training, our method shows superior generalization ability on clinical Splveic data. Our future work will focus on more complex scenarios in real situations, as well as further reducing workload in actual production in an unsupervised/self-supervised manner.

\begin{acks}
To Robert, for the bagels and explaining CMYK and color spaces.
\end{acks}

\clearpage
\bibliographystyle{ACM-Reference-Format}
\bibliography{Refs_RetinexFlow}

\appendix

\end{document}